\documentclass[aps,prb,twocolumn,superscriptaddress,amsmath,amssymb]{revtex4-2}
\usepackage{graphicx} 
\usepackage{mathrsfs}
\usepackage{gensymb,babel}
\usepackage{xr}
\usepackage{cleveref}
\usepackage[T1]{fontenc}

\begin{document}

\title{Ferroelectric to incommensurate fluctuations crossover in PbHfO$_3$-PbSnO$_3$}

\author{Maria A. Kniazeva}
\email[]{kniazeva.maria225@yandex.ru}
\author{Alexander E. Ganzha}
\affiliation{Peter the Great Saint-Petersubrg Polytechnic University, 29 Politekhnicheskaya, 195251, St.-Petersburg, Russia} 
\author{Irena Jankowska-Sumara}
\affiliation{Institute of Physics, Pedagogical University of Cracow, Podchorazych 2, 30-084 Krakow, Poland}
\author{Marek Pasciak}
\affiliation{Institute of Physics, The Academy of Sciences of the Czech Republic, Prague, Czech Republic}


\author{Andrzej Majchrowski}
\affiliation{Institute of Applied Physics, Military University of Technology, ul. Gen. W. Urbanowicza 2, 00-908 Warszawa, Poland}
\author{Alexey V. Filimonov}
\affiliation{Alferov University, 8-3-A Khlopina, 194021, St.-Petersburg, Russia}
\affiliation{Peter the Great Saint-Petersubrg Polytechnic University, 29 Politekhnicheskaya, 195251, St.-Petersburg, Russia} 
\author{Andrey I. Rudskoy}
\affiliation{Peter the Great Saint-Petersubrg Polytechnic University, 29 Politekhnicheskaya, 195251, St.-Petersburg, Russia}
\author{Krystian Roleder}
\affiliation{Institute of Physics, University of Silesia, 75 Pułku Piechoty 1, 41-500 Chorzów, Poland}
\author{Roman G. Burkovsky}
\email[]{roman.burkovsky@gmail.com}
\affiliation{Peter the Great Saint-Petersubrg Polytechnic University, 29 Politekhnicheskaya, 195251, St.-Petersburg, Russia} 

\begin{abstract}

Perovskite solid solutions PbHfO$_3$-PbSnO$_3$ offer valuable opportunities for studying the formation mechsnisms of incommensurate phases, owing to the presence of an intermediate (between cubic and incommensurate) phase, which is stabilized in PbHfO$_3$ upon PbSnO$_3$ admixture. Here, x-ray diffuse scattering signal is used to quantify the evolution of susceptibilities related to different modes of distortion (ferroelectric, incommensurate, antiferrodistortive) as a function of temperature and the results are critically compared to the predictions of a minimal symmetry-based Landau-like model with two coupled order parameters (ferroelectric and antiferrodistortive) and incommensurate order parameter being interpreted as inhomogeneous polarization. 
Experimentally, we observe a Curie-Weiss-like linear dependence of ferroelectric stiffness (inverse of susceptibility related to homogeneous polarization fluctuations) in the cubic phase down to about 50 K above the transition to the intermediate phase, where this dependence nearly saturates. Upon cooling down to the intermediate phase, the maximum of susceptibility shifts gradually to the non-zero wavevector, where another Curie-Weiss-like linear stiffness trend is established, but with respect to the incommensurate order parameter. Symmetry of diffuse scattering distributions indicates an orthorhombic symmetry of the intermediate phase.
A notable temperature dependence of the constant that describes the energy of polarization inhomogeneities is observed experimentally, which is in disagreement with the model expectations. Specifics of this dependence suggest the presence of a nearly temperature-independent characteristic length scale for inhomogeneities across several phases.  
Other differences with the model suggest that the incommensurate order parameter cannot be straightforwardly identified with weakly-inhomogeneous ferroelectric polarization.

\end{abstract}

\maketitle

\section{Introduction}

Anti-polar ordering of cations in perovskite crystals attracts long time attention because of intrinsic scientific interest \cite{Kittel1951,Lawless1967,Bellaiche2013} and due to promising applications of antiferroelectrics \cite{Rabe2013,Wei2014,Pesic2016make,Park2016,Xu2017}. These materials are a fruitful test ground for electron microscopy studies \cite{Viehland1994,Cai2003}, scattering experiments \cite{Tagantsev2013,Zhang2015neutron,Burkovsky2017,Bosak2020} and theories \cite{Kittel1951,Haun1989,Bussmann2013,Bellaiche2013,Tagantsev2013,Iniguez2014,Hlinka2014,Patel2016,Toledano2016,Burkovsky2018,Xu2019order,Schranz2020contributions}. A significant share of theoretical attention is given to interacting order parameters in PbZrO$_3$-like antiferroelectrics and to the mechanisms of incommensurate phases formation. This paper provides an experimental material related to those questions by reporting a diffuse scattering study of Pb(Hf$_{0.7}$Sn$_{0.3}$)O$_3$. This crystal has an intermediate phase, where oxygen octahedral tilts are already developed, while the incommensurate order parameter exists only in the form of fluctuations. Studying those fluctuations by diffuse scattering helps identifying the important order parameter interaction terms in Landau-like phenomenological model and register the incommensurate soft mode that has been experimentally elusive in crystals of this family.


\subsection{Models of antiferroelectrics}
Complexity of antiferroelectrics resulted in a rather large number of different approaches and perspectives in their modeling. To help the reader navigate that, a brief systematization follows.

First model of antiferroelectrics is due to Kittel \cite{Kittel1951}, who reduced an (at that time -- abstract) antiferroelectric to two inter-penetrating lattices with anti-parallel polarizations. The most studied real antiferroelectric, PbZrO$_3$, is much more complex than that, in part -- due to the more complex dipole ordering $\uparrow\uparrow\downarrow\downarrow$ \cite{Whatmore1979}, due to the possibility of alternative incommensurate orderings \cite{Viehland1994} and due to the presence of anti-phase octahedral tilts as a second type of distortion \cite{Whatmore1979}, usually coexisting with the anti-polar ordering of cations. 

Modern models fall, roughly, into three categories. First -- \textit{ab-initio} analysis \cite{Singh1995,Wagmare1997,Ghosez1999} of energy landscape at zero Kelvin, sometimes extended to modeling of finite temperature behavior \textit{via} molecular dynamics \cite{Pasciak2015,Mani2015,Xu2019order}. Usually this extension needs a parametrization of energy in terms of local ionic displacements around high-symmetry positions (often referred to as second-principles approach). Alternatively, the energy can be parametrized not in local coordinates, but in normal-mode coordinates \cite{Iniguez2014,Xu2019order}, which is not suitable for molecular dynamics, but allows some more immediately comprehensible insights into the energetics of the crystal, in particular -- into coupling of different degrees of freedom. 

Second category -- semi-empirical atomistic models -- is more multi-faceted. In contrast to \textit{ab-initio}, these models contain assumptions and phenomenological parameters, which makes them more flexible for exploring narrow and specific effects. For example, the simplified lattice-dynamics model of Ref. \cite{Bussmann2013} suggests an atomistic picture behind temperature-dependent avoided crossing of acoustic and optic phonon branches, model of Ref. \cite{Bellaiche2013} points to universal collaborative coupling between octahedral tilts and antiferroelectricity, model of Ref. \cite{Hlinka2014} suggests a mechanism of dipole wave commensuralization in the presence of flat and soft polarization branch. More recently, Ref. \cite{Patel2016} explores the link between bi-linear coupling and incommensurate structures, Ref. \cite{Burkovsky2018} shows how such structures can arise due to the dipole-dipole interactions, in analogy with polarization catastrophe in ferroelectrics.

Third category is Landau-like models, similar to the one of Devonshire for BaTiO$_3$ \cite{Devonshire1949}. These models do not account for the specifics of atomic structure and short-range interactions (in contrast to long-range electric and elastic effects \cite{Marton2006simulation}), but parametrize macroscopic thermodynamic potential based on symmetry arguments. In contrast to Landau theory of second-order phase transitions \cite{Landau2013statistical}, Landau-like models do not claim exactness in any applicability domain \cite{Levanyuk2020}, but often describe the experimental observations adequately. 

For antiferroelectric perovskites, such an approach was applied by Haun \textit{et al.} \cite{Haun1989}, who have reproduced electric and elastic behavior in PbZrO$_3$ by a model with ferroelectric and antiferroelectric order parameters coupled to each other and to the stress. Those parameters were considered homogeneous. Toledano and Khalyavin considered a similar model, but with coupling also to anti-phase octahedral tilts \cite{Toledano2019}. Tagantsev \textit{et al.} \cite{Tagantsev2013} proposed a Landau-like model for the same crystal where anti-polar displacements correspond not to the independently-defined antiferroelectric order parameter (as in Haun's work), but to inhomogeneous polarization. Additionally, that model accounts for inhomogeneous strain, acousto-optic (flexoelectric) coupling, Umklapp interactions and bi-quadratic coupling between acousto-optic waves and anti-phase octahedral tilts. 

Presently there is no consensus on why the PbZrO$_3$-like crystals tend to be incommensurate instead of ferroelectric. Tagantsev \textit{et al.}'s proposition \cite{Tagantsev2013} that it is due to flexoelectric interaction is in line with a broader discussion on the possible origin of incommensuration in dielectrics \cite{BlincLevanyuk,Heine1984,Saint1999role} due to the presence of an invariant term of the form $P\partial u/\partial x - u\partial P/\partial x$, where $P$ is the polarization and $u$ is the strain or another quantity with appropriate transformational properties. A point of confusion in that broader discussion has been summarized by Levanyuk \cite{BlincLevanyuk} as \textit{``In the literature one may find the assertion that it is this coupling, which is a general reason for [incommensurate] phase formation. Such a statement would be meaningful only if one could be sure that the nonrenormalized coefficient [of polarization correlation energy] is positive. But we hardly know confidently if it is the case ...''}. There are few microscopic models in the literature \cite{Bussmann2014,Patel2016,Burkovsky2018} that suggest a way of obtaining the negative \textit{nonrenormalized} coefficient above due to particular atomic-level specifics of perovskites. This ambiguity seems presently distant from the resolution.

\subsection{Motivation}
There is a considerable experience of simulating polarization fluctuations, as seen by scattering, in the cubic phase of perovskites  \cite{Andronikova2016Critical,Burkovsky2019}, where there are no other order parameters with which the polarization could interact. When the symmetry becomes lower than cubic, the goal becomes more complex \cite{Burkovsky2017}. Positive side of that complication is, however, an emerging possibility of studying the energetics of order parameter interactions in low-symmetry phases. 

A natural test ground for this could be the high-pressure intermediate phase in PbZrO$_3$ \cite{samara1970pressure, Burkovsky2017} and PbHfO$_3$ \cite{samara1970pressure, Knyazeva2019} that becomes stable between the cubic and the incommensurate phases almost immediately on raising pressure above ambient. This phase is of lower-than-cubic symmetry, has an additional order parameter (octahedral tilts, as Ref. \cite{Knyazeva2019} suggests), apparently no long-range ordered displacements in cation subsystem, but strong fluctuations of inhomogeneous polarization, as seen by diffuse scattering. It would be seen as a good model object, but studying it in temperature domain is difficult, since pressure complicates such experiments considerably. A possible strategy is to study a similar phase, but created, somehow, without pressure, for example – by chemical modifications.

\subsection{Tin-doped PbZrO$_3$ and PbHfO$_3$}
It is instructive to consider tin-doped PbZrO$_3$ and PbHfO$_3$. In contrast to titanium doping, which leads to ferroelectric PZT \cite{Jaffe1971}, doping by tin leads in a completely different direction \cite{Sumara2017}. Tentatively, the fully-filled $d$-shell of tin prevents it form acting like titanium, but results in effect similar to that of pressure due to ionic radius difference. As a result of tin doping the intermediate phase forms, appearing highly similar to the one(s) under pressure.

In the parent compound of PbHfO$_3$-PbSnO$_3$, pure PbHfO$_3$, two phase transitions occur on varying temperature: at about $163^{\circ}$ $C$, the low-temperature antiferroelectric (AFE1) phase \cite{Corker1998}, transforms to the incommensurately modulated AFE2 phase \cite{Fujishita2018structural,Burkovsky2019,Bosak2020}, and at about $215^{\circ}$ $C$ the cubic phase forms. Curiously, cubic$\rightarrow$AFE2 transition is a triggered incommensurate transition, which is controlled not by an incommensurate soft mode, as in most of the incommensurate dielectrics \cite{BlincLevanyuk}, but by the softening of the antiferrodistortive (AFD) mode \cite{Burkovsky2019}. Triggered transition scenario is, seemingly, suppressed on doping by tin, because an intermediate phase between the cubic and the AFE2 phases appears, allowing the distortions related to the AFD and incommensurate order parameters to form at different temperatures \cite{jankowska2020local}. There is no sharp drop in the dielectric constant in the intermediate phase \cite{Sumara2017} compared to the cubic phase, which suggests the intermediate phase to be non-polar.

\subsection{This work}
This paper reports on a study of the intermediate phase and related transitions by diffuse scattering in PbHfO$_3$-PbSnO$_3$ solid solution with 30$\%$ concentration of tin atoms -- a nearly critical concentration in terms of solubility.

Primary observation of this work is a strongly temperature-dependent incommensurate diffuse intensity maximum in the intermediate phase, which emerges gradually from zone-center maximum in cubic phase. To put that onto a quantitative basis, we propose recalculating intensities to stiffness with respect to distortions, in which case the situation becomes more understandable. This leads to further notes, such as unusual saturation of dielectric stiffness in cubic phase and surprisingly linear (Curie-Weiss) trend for incommensurate stiffness going to zero at the point of incommensurate transition. 

Particularly interesting is the behavior of correlation energy, which characterizes how much does the crystal resist inhomogeneities in polarization. In contrast to what is expected for normal ferroelectric crystals, the one under study has a notably temperature-dependent correlation energy, not only in the intermediate phase, where, as we suggest, interaction between its order parameter (tilts) and polarization gradient could explain that, but also in the cubic phase, where it occurs spontaneously. This temperature dependence of correlation energy suggests a possibility of an unusual largely temperature-independent characteristic length scale for inhomogeneities.

The findings help advancing the experimental and interpretation methodology, as well as add some particular physical insight, especially on the interaction of octahedral tilts and dipoles arrangement.

\section{Methods}

\subsection{Idea}
Stiffness with respect to particular normal modes of ionic displacements, $\alpha$, can be studied by diffuse scattering. This is possible because of the following relationship between stiffness and fluctuations:
\begin{equation} \label{eq_fluctuations}
    \langle P^2 \rangle = \frac{T}{\alpha},
\end{equation}
where $\langle P^2 \rangle$ is ensemble averaged square of normal mode amplitude, $P$. This follows from equipartition theorem, stating that the average thermal energy deposited into a normal mode in equilibrium, $\alpha \langle P^2 \rangle/2$ equals to $T/2$. Diffuse scattering signal due to that normal mode is proportional to $\langle P^2 \rangle$, so $\alpha$ can be extracted upon taking into account the corresponding structure factor, provided that the relevant signal is separated from other scattering.

This stiffness is the same as the one determining linear response to a conjugate field, $E$ as
\begin{equation} \label{eq_linear_response}
    P = \alpha^{-1} E.
\end{equation}

If $P$ is polarization and $E$ is the electric field (any order parameter and conjugate field can be substituted here), then $\alpha^{-1}$ should be dielectric susceptibility, as measured in capacitors. This links scattering measurements, sensing temperature-induced $\langle P^2 \rangle$, to linear response measurements, sensing $P$ as a function of $E$.


\subsection{Experimental}
Single crystals of PbHfO$_3$-PbSnO$_3$ have been grown by spontaneous crystallization method in Military Institute of Technoloby, Warsaw, Poland, as described in Ref. \cite{Sumara2017}.  
Experiments on diffraction and diffuse scattering have been performed at the European Synchrotron Radiation Facility (ESRF), at the side station of ID28 beamline. For measurements we have used a needle-like piece with cross-section about 100x100 microns, mounted on quartz capillary and heated in the flow of hot air due to the heat blower. The measurements were done in a cooling cycle. Tight focusing of the beam  (about 40 microns) and relatively large size of the sample, as compared to the characteristic attenuation length of about 10.5 microns (wavelength of $\lambda$= 0.6968 ${\AA}$), induced a considerable inhomogeneity in the overall signal distribution in reciprocal space, prompting us to analyze the data corresponding to a sufficiently homogeneous part of it. The exposure was 0.2 - 0.5 secs per 0.1$^{\circ}$ angular step. Data treatment has been carried out using CrysAlis Pro program and custom-built Matlab codes.

\subsection{Interpreting and modeling the signal}

Diffuse scattering (DS) is a scattering corresponding to wavevector transfers, $\vec{Q}$, that are outside the reciprocal lattice points, $\vec{\tau}$. It is due to disorder in the arrangement of scattering centers, which in the case of crystals under study corresponds, mainly, to partially disordered, but still correlated displacements of ions from their nominal positions in the structure. 

DS from fluctuations of a particular normal mode (order parameter) is proportional to susceptibility, according to the formula \cite{Bruce1981}:
\begin{equation}
{I}(\vec{Q},T) = T\cdot| F_{\text{DS}} (\vec{Q}) | ^ 2\cdot\chi(\vec{q},T),
\label{susc}
\end{equation}
%
where $T$ is the temperature in energy units, $F_{\text{DS}}$ is the structure factor for diffuse scattering, $\chi=\alpha^{-1}$ is the susceptibility with respect to that normal mode. The structure factor $F_{\text{DS}}$ is computed similary to the inelastic structure factor for scattering by phonons \cite{Bruce1981,Dorner1982}. Relatively simple cases of the present paper (transverse polarization waves and tilts of octahedra around a particular axis) can be treated with a simplified form for the structure factor
\begin{equation}
F_{\text{DS}} \sim (\vec{Q}\vec{u}_{\vec{q}}),
\label{Fds}
\end{equation}
where $\vec{u}_{\vec{q}}$ is the Fourier component of displacements distribution, corresponding to reduced wavevector $\vec{q}=\vec{Q}-\vec{\tau}$. This simplification is obtained (see Appendix) by neglecting the interference of scattering centers within the cell, but keeping the essential dependence on the directions of displacements. 

\begin{figure*}
\includegraphics[width = .7\textwidth]{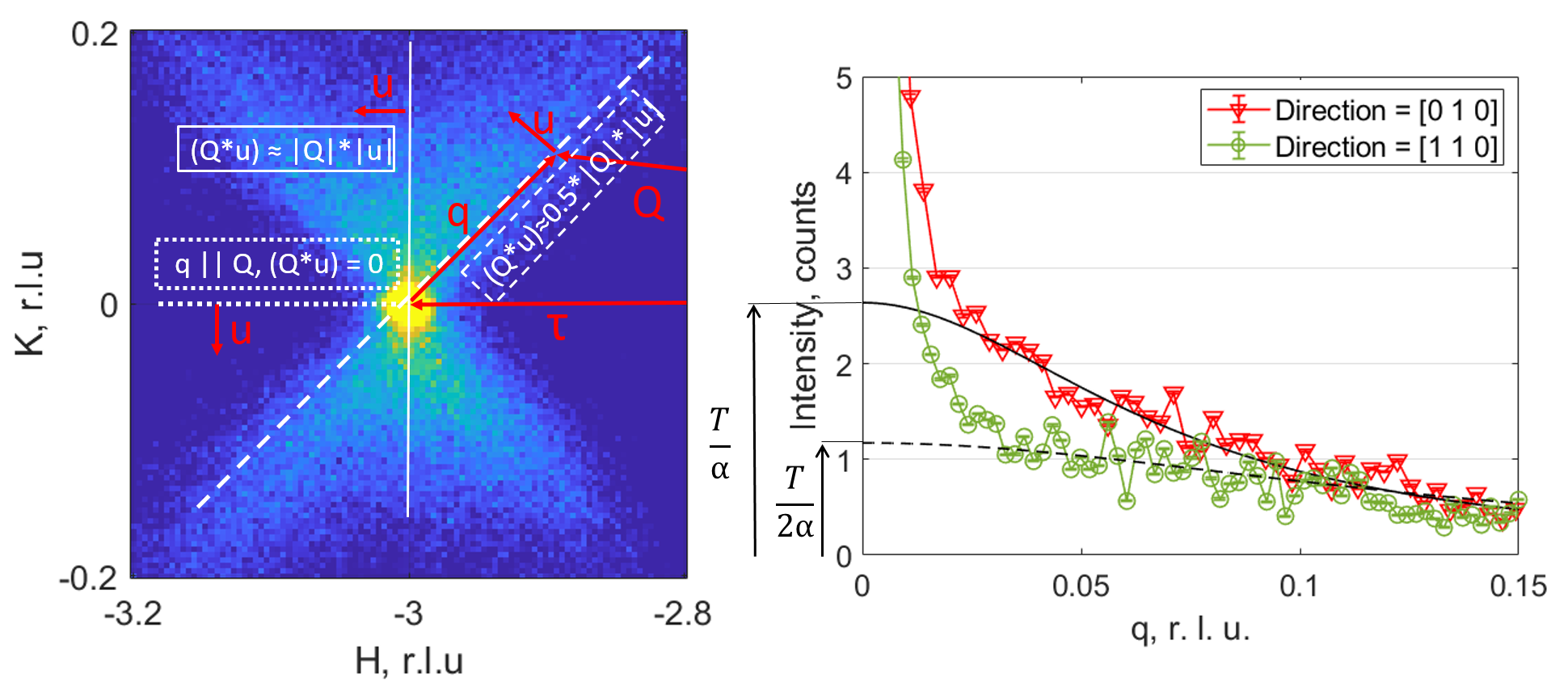}
\caption{\label{DS in PE}
Schema of determining the stiffness with respect to transverse polarization waves in the cubic phase. Intensity along $\tau$ corresponds to the background (which is substracted from profiles), because of zero structure factor for transverse waves. Stiffness at zero wavevector is obtained by fitting finite-wavevector signal using Lorentzian line shape. Such fits along different directions provide mutually consistent estimates of $\alpha(\vec{q}=0)$.
}
\end{figure*}

Stiffness with respect to a particular order parameter, when considered as a function of wavevector, should be parabolic near $q=0$:
\begin{equation}
\alpha(\vec{q}) = \alpha + \sum_{i,j} D_{i,j} q_i q_j,
\end{equation}
which reflects that homogeneous order parameter should be energetically less expensive than an inhomogeneous one (as long as there is no Lifshitz invariant, which holds for cubic perovskite structure). Therefore, when considering an arbitrarily oriented line in reciprocal space going through a reciprocal space node, one should expect Lorentzian-shaped intensity profile
\begin{equation}
I(\vec{Q}) \sim \frac{T | F_{\text{DS}} (\vec{Q}) | ^ 2}{\alpha + D_2 |\vec{q}|^2},
\label{Chi}
\end{equation}
where $D_2$ is a constant of spatial correlation energy for that order parameter along that direction. In the case of polarization waves, one should count $\vec{q}$ as originating at the cubic $\Gamma$ points, while in the case of oxygen octahedral tilts, one should treat it as originating from the $R$ points.
Basically, the data treatment proceeds by experimentally quantifying the parameters in formula (\ref{Chi}) and, in the less trivial cases -- beyond it. An annotated example of extracting the dielectric stiffness from diffuse scattering in cubic phase is shown in Fig. \ref{DS in PE}. 

Note that this approach can be justified only near the parent Bragg reflections with small structure factor. Small structure factor for Bragg scattering implies small structure factor for scattering by long-wavevlength acoustic waves. This allows sensing mostly distortions of the cells, while the shifts of the cells as a whole are discarded to a large degree. For that purpose, this work uses (-3, 0, 0) Bragg peak, which has a relatively small structure factor, about 1/10 of that for the strong (2, 0, 0) peak.

The diffuse scattering data of the present work are treated as if the scattering originated from fluctuations of inhomogeneous polarization only. In fact, these waves may be \textit{``renormalized''} due to flexoelectricity or other similar-in-form interactions (see Introduction) and contain, therefore, a mixture of polarization and some other property. This possibility is ignored in direct data modeling of the present work, mainly because the data do not allow distinguishing one possible renormalization scenario from the other. It is assumed that the present data treatment results could be reassessed upon obtaining a more specific understanding of renormalization in the future.


Order parameter magnitudes are seen in diffraction by intensities of superstructure reflections. When distortions are small, these intensities are proportional to the square of the distortion magnitude \cite{Bruce1981}. So, if the transition is of second order, one expects nearly linear temperature dependence of intensities, $I \sim (T_0-T)$, without discontinuities. In the first-order transition case, one expects a jump at the transition temperature.

\section{Results}

\subsection{Two phase transitions}

\begin{figure}
\includegraphics[width=\columnwidth]{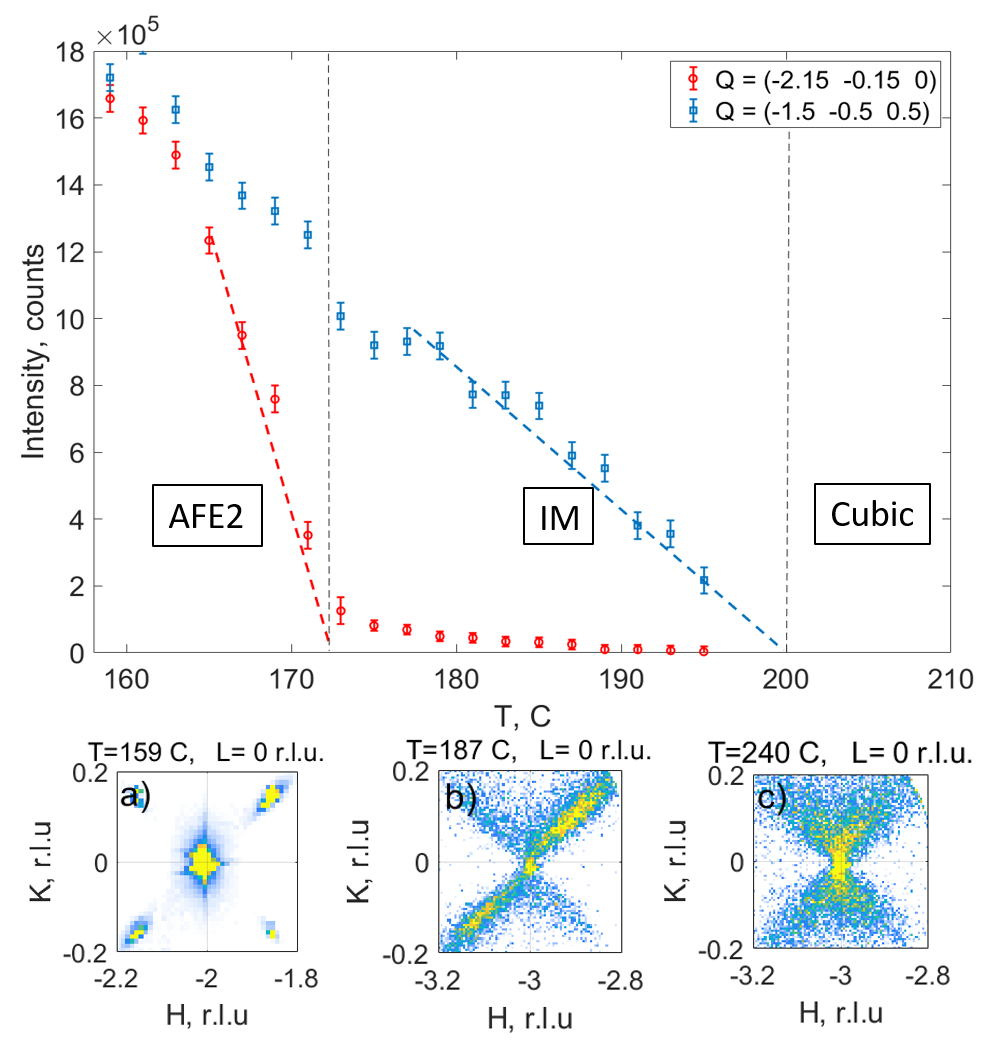}
\caption{\label{Phases}
Three phases of Pb[Hf$_{0.7}$Sn$_{0.3}$]O$_3$: incommensurate AFE2 (a),  intermediate IM (b) and cubic (c). Squares -- intensity at $R$-points, circles – intensity of $\Sigma_F$-reflections. Non-zero values above the IM-AFE2 transition are due to intensity of diffuse scattering at the incommensurate positions. Dashed lines show the transition temperatures.
}
\end{figure}

The range of temperatures that have been studied contains three different phases: incommensurate phase (AFE2), intermediate phase (IM), and the cubic phase. Their characteristic scattering patterns are shown in Fig. \ref{Phases}. The low-symmetry AFE2 phase shows incommensurate reflections at  $\Sigma_F$ ($h \pm \xi_F $, $k \pm \xi_F$, $l$) points and zone-boundary reflections at $R$ ($h +0.5 $, $k +0.5$, $l+0.5$) points, which is consistent with structural reports on the same phase in pure PbHfO$_3$ \cite{Fujishita2018structural,Bosak2020}.
In the higher-temperature IM phase, the sharp reflections at $\Sigma_F$ disappear, while diffuse scattering maxima about the same positions emerge. 

As to the $R$-point reflections in the IM phase, they demonstrate a particular systematics. All the inspected (few tens) reflections at the $R$-points with symmetric indices, $|H|=|K|=|L|$, are by about an order of magnitude less intense than some of the reflections with non-symmetric indices. From this one may suggest that the $R$-point reflections are mostly due to the octahedral tilts. If they were totally due to the tilts, the symmetric reflections would be expected totally absent in the small-displacement limit \cite{Glazer1975}. The experimental presence of small intensities at symmetric $R$-points can be attributed to the minor presence of an additional distortion mode (such as anti-phase Pb shifts) or to the violation of small-displacement limit in the experiment. Tentatively, the octahedral tilt pattern is the same as in the AFE1 and AFE2 phases ($a^-a^-c^0$), but with smaller amplitude. 

In the cubic phase, there are no superstructural reflections, and diffuse scattering has a butterfly-like shape with a maximum at the center of the Brillouin zone. 

From scattering, both transitions appear to be close to the second order. Intensities of the corresponding superstructural reflections increase linearly directly near the transition temperatures. 
It seems most natural to estimate the transition temperatures by extrapolating the linear sections to the temperature axis - $T_\text{AFD} \approx 200^{\circ}$ $C$ and $T_\text{IC} \approx 172^{\circ}$ $C$.

\subsection{Antiferrodistortive mode critical dynamics}

Since the transitions appear close to the second order, a critical increase of respective susceptibilities should occur on approaching the transition temperatures. That should manifest in diffuse scattering intensity increase at and near the corresponding points of the Brillouin zone, which is, for AFD tilting -- the $R$-point.

\begin{figure*}
\includegraphics[width=0.7\textwidth]{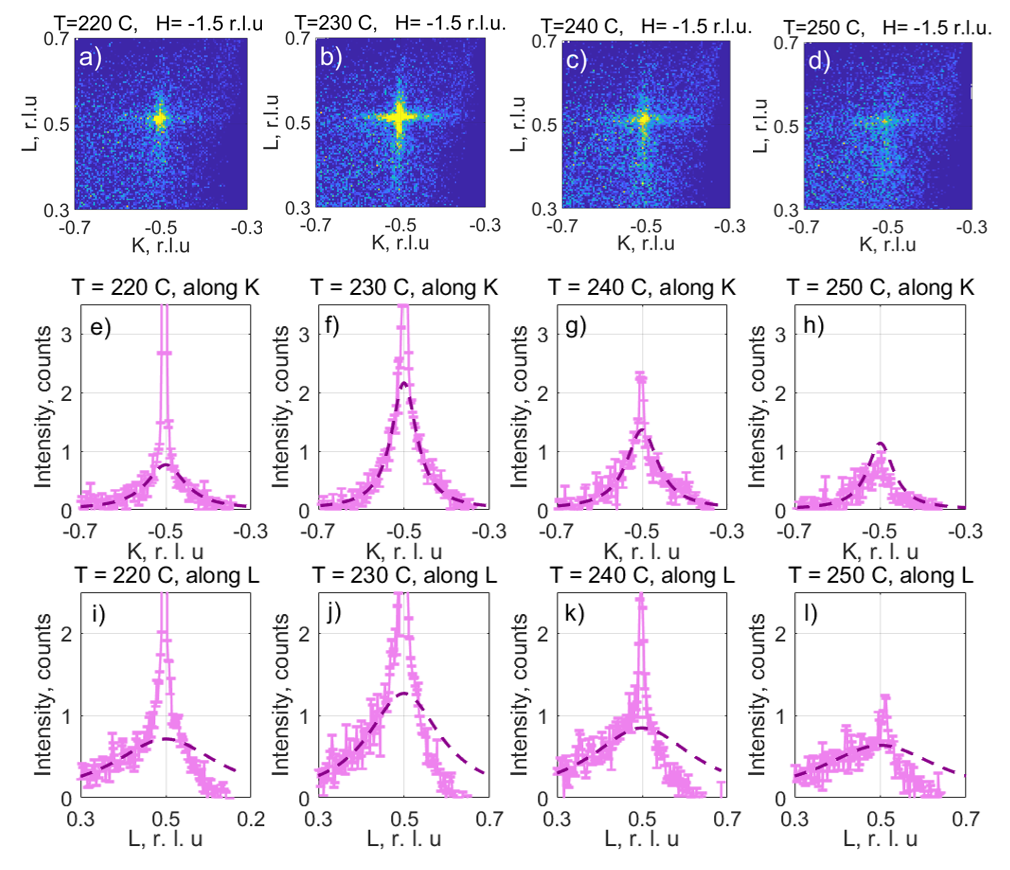}
\caption{\label{Crosses}
Diffuse scattering distribution in the vicinity of the (-1.5 -0.5 0.5) reflection in the cubic phase. Profiles (e-l) are constructed by integrating the signal in cylinders with a radius of 0.014 r.l.u. Dashed lines are the fit by Lorentzian $I = \frac{T}{\alpha + D_2(\vec{q} - \vec{q_0})^2}$. Asymmetry of the $L$-direction profile is  likely due  to the experimental artifact: in some places of the reciprocal space there are strong intensity gradients due to the relatively large size of the sample in comparison with the characteristic absorption length of radiation in it.
}
\end{figure*}

\begin{figure}
\includegraphics[width = .8\columnwidth]{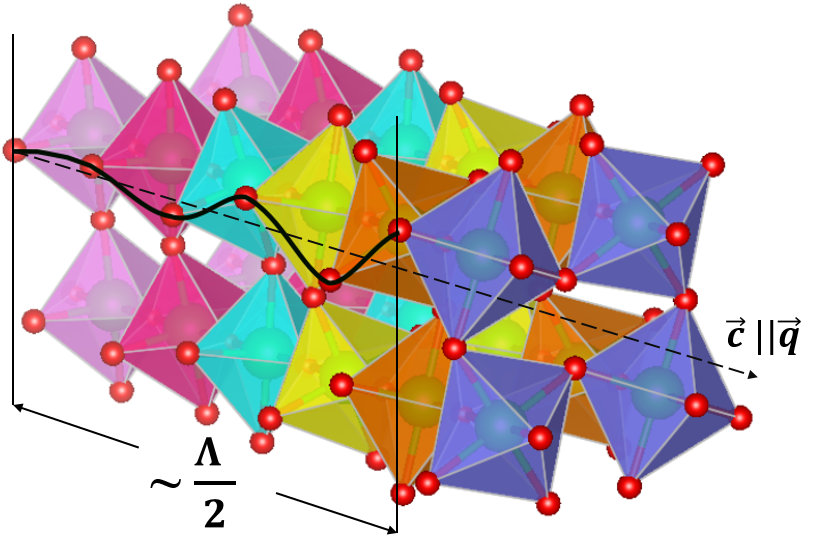}
\caption{\label{O_tilts}
A real-space view of an AFD fluctuation responsible for diffuse scattering rods near $R$-points. Octahedra experience anti-phase tilts around the axis that coincides with the direction of wavevector, i.e. direction in which the pattern slowly changes in real space. In the present case, this change is assumed as just the decay of tilt magnitude over the characteristic correlation length, $\Lambda$. Note that this pattern of short-range tilt correlations is similar to $a^0a^0c^-$, which is different from the tilt pattern in the antiferroelectric, incommensurate and (presumably) IM phases, which is $a^-a^-c^0$.
}
\end{figure}

Diffuse cross-shaped distributions (Fig. \ref{Crosses}) are formed by the intersecting diffuse scattering rods oriented along the $H$, $K$ and $L$ directions. There are very weak or no rods for which $\vec{q}$ = $\vec{Q}-\vec{Q}_R$ is approximately parallel to $\vec{Q}$. Consequently, the displacements from which the scattering stems are perpendicular to $\vec{q}$. Taking into account that we consider antiphase tilts, the tilts are around the direction of $\vec{q}$. Visually, such an inhomogeneous fluctuation in tilt subsystem can be imagined as in Fig. \ref{O_tilts}.
As the cubic$\rightarrow$IM transition is approached, the intensity of the rods increases, as can be seen from the profiles. Unexpectedly, the intensity of the rods drops sharply at $T = 220^{\circ}$ $C$, which is still significantly higher than the transition temperature obtained from the extrapolation of the amplitude of the tilting ($T = 200^{\circ}$ $C$).

\begin{figure}
\includegraphics[width = \columnwidth]{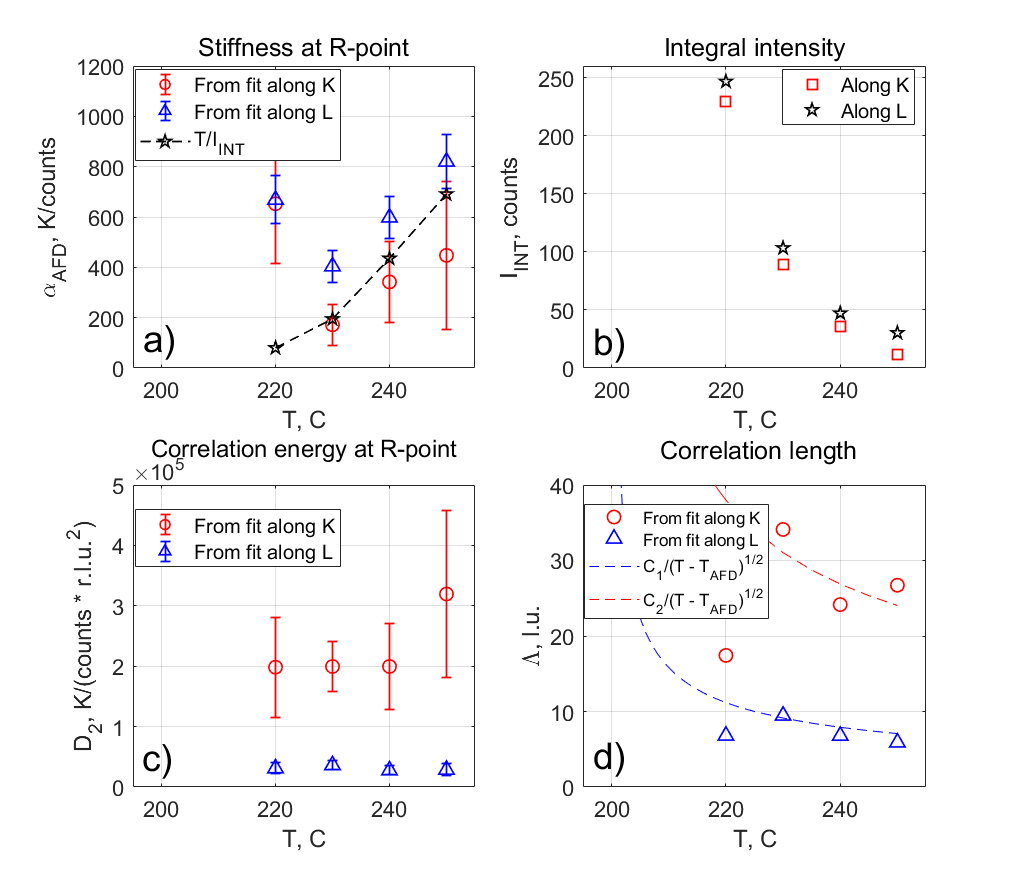}
\caption{\label{Stiffness and Length}
Diffuse scattering parameters in the vicinity of the $R$-point (-1.5 -0.5 0.5): (a) AFD  stiffness $\alpha_\text{AFD}$ determined by independent fits along the $K$ and $L$ axes. We think that the true stiffness, which does not depend on the direction in cubic phase, is average between them. Stars are the stiffness recalculated from integral intensity of the profiles along $L$. (b) Integral intensity for the corresponding directions. (c) Correlation energy constant $D_2$. (d) Correlation lengths of displacements recalculated from stiffness and correlation energy constant. Dashed lines are theoretical expectations under the assumption $D_2$ = const and $\alpha_\text{AFD} = A (T - T_\text{AFD})$.
}
\end{figure}

In the language of stiffness, this corresponds first to its gradual decrease on cooling and then -- in an abrupt  increase at $T = 220^{\circ}$ $C$ [Fig. \ref{Stiffness and Length} (a)].

If one inspects not the extrapolation of DS intensity to the $R$-point, but  the integrated intensity in its vicinity, then the picture becomes different [Fig. \ref{Stiffness and Length} (b)]. Integral intensity grows without unexpected features. The reason is that at temperature where diffuse scattering is suppressed ($T = 220^{\circ}$ $C$) the sharp peak at the $R$-point itself increases strongly (it is still much weaker than in the IM phase), and the total intensity nevertheless increases. Upon recalculating this integral intensity value into a stiffness, the resulting effective stiffness behaves as expected -- decreases on approaching the transition [dashed line in Fig. \ref{Stiffness and Length} (a)].

It is natural to explore how it looks from the perspective of tilt-tilt correlation length [Fig. \ref{Stiffness and Length} (d)]. The formula for this quantity, $\Lambda = \sqrt{D_2/\alpha_{\text{AFD}}}$, is extracted from denominator of Eq. \eqref{Chi} as the combination with dimension of length. At temperatures that are above diffuse intensity drop, the length grows on cooling, agreeing with the theoretical expectations, which are shown by dashed lines [Fig. \ref{Stiffness and Length} (d)]. At the moment of diffuse intensity drop, the length also decreases sharply. It turns out that the correlation length associated with normal diffuse scattering does not grow significantly higher than 20-30 cells. On the other hand, the continued growth of the integral intensity, and, in particular, the narrow Bragg-like component directly at the $R$-point, suggests that the degree of ordering of the tilting still grows further. It is likely that for an adequate assessment of the stiffness and correlation length it is necessary to take into account both the normal diffuse signal (Lorentzian-like) and the narrow component directly at the $R$-point. At the moment, it is not clear how this should be made most consistently (we used intensity integration as the simplest approach). Parallels can be seen with phonon spectroscopy, where energy integration with taking into account also the central peak, appears necessary to estimate the ``true'' stiffness \cite{Axe1973,Burkovsky2019}.

An alternative suggestion could be that maybe the sample at $T = 220^{\circ}$ $C$ is not fully homogeneous and some sub-volumes of it are already in the ordered (IM) phase, while the rest of the crystal is in cubic phase. The Bragg-like component is, in this case, just a normal Bragg reflection from those ordered regions. In the case of SrTiO$_3$ a similar question arose (see Ref. \cite{Bruce1981}, section III.2.1, for review) and was the subject of targeted studies \cite{Cowley1978quasi}, which were aiming at elucidating the possibility of higher transition temperature in near-surface layers, but did not confirm that.

\subsection{Unrealized transition to the polar phase}

In addition to an increase in the AFD susceptibility, an increase in the dielectric susceptibility is also expected in the cubic phase, according to the dielectric studies of Pb(Hf$_{0.7}$Sn$_{0.3}$)O$_3$ crystals \cite{Sumara2017}.

\begin{figure}
\includegraphics[width = \columnwidth]{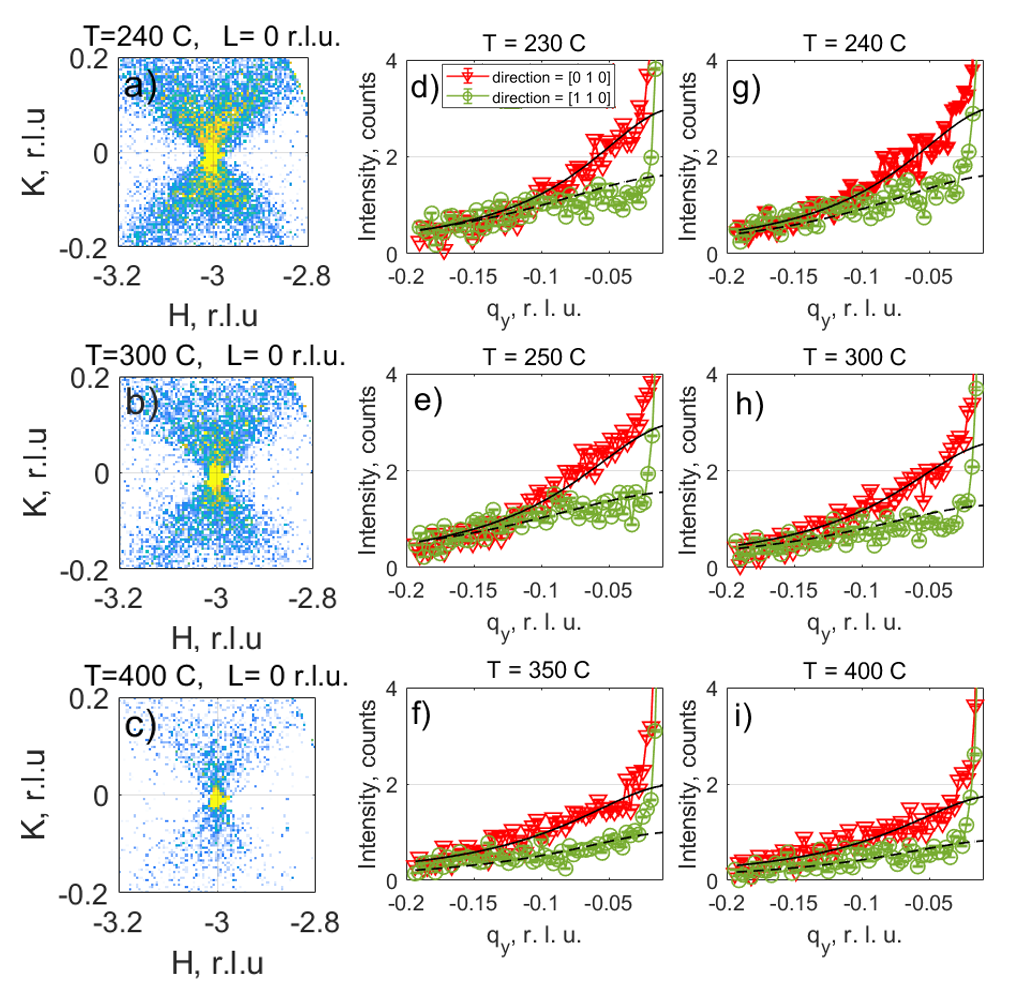}
\caption{\label{DS_PE}
Distribution of diffuse scattering in the cubic phase. Two-dimensional distributions (cuts througth $Q$ = (-3 0 0) by $H$ $K$ $0$ plane) are shown using same color scale. 
The profiles correspond to [1 1 0] (circles) and [0 1 0] (triangles) directions, background  subtracted. Solid and dashed lines are fit by Lorentzian $I = \frac{T}{\alpha + D_2(\vec{q} - \vec{q_0})^2}$.
}

\end{figure}

\begin{figure}
\includegraphics[scale=0.35]{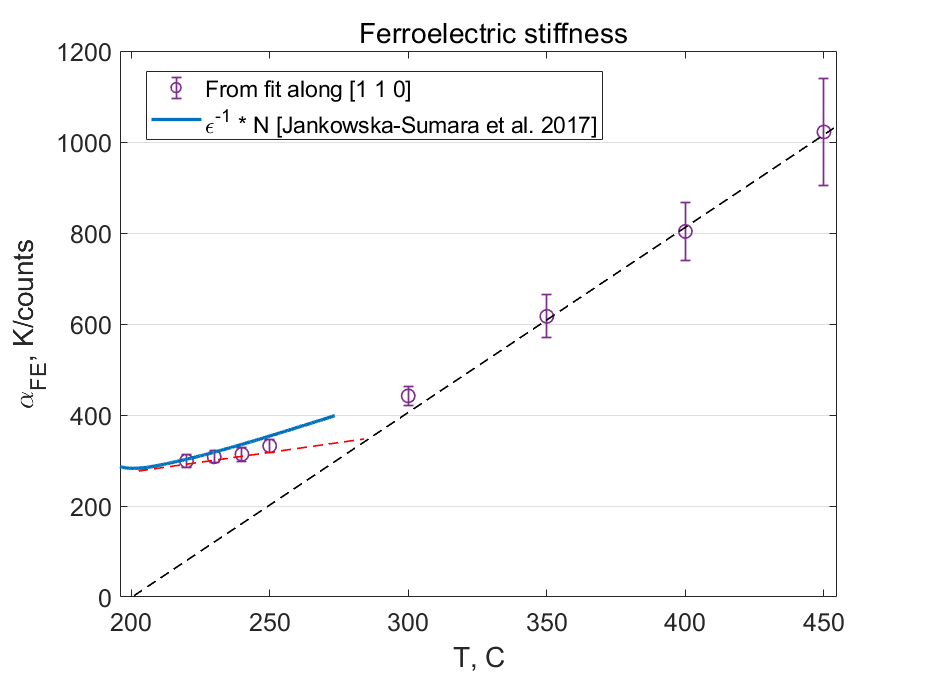}
\caption{\label{alpha_FE}
The temperature dependence of the dielectric stiffness, determined by the analysis of diffuse scattering in the cubic phase compared to dielectric measurements \cite{Sumara2017}. For a correct comparison, the moment of the phase transition of the dielectric data was shifted in temperature to the value T = $200^{\circ}$ $C$, at which a phase transition is observed in the current study. 
}

\end{figure}

The  stiffness at $\vec{q}=0$, denoted as $\alpha_\text{FE}$ and determined by the fitting of diffuse scattering profiles (Fig. \ref{DS_PE}), decreases on cooling (Fig. \ref{alpha_FE}). At high temperatures ($300 - 450^{\circ}$ $C$) this fits to Curie-Weiss law with a critical temperature $T_\text{FE} \approx 200^{\circ}$ $C$, which coincides unexpectedly well with the AFD transition temperature. At lower temperatures ($220 - 250^{\circ}$ $C$) the stiffness decrease slows down, nearly saturates. This lower-temperature saturation seems qualitatively compatible with dielectric data of Ref. \cite{Sumara2017}, which is co-plotted. 
The absence of divergence of susceptibility at zone center agrees well with the absence of ferroelectric transition (AFD transition occurs instead). 

\subsection{Incommensurate mode critical dynamics}

\begin{figure}
\includegraphics[width = \columnwidth]{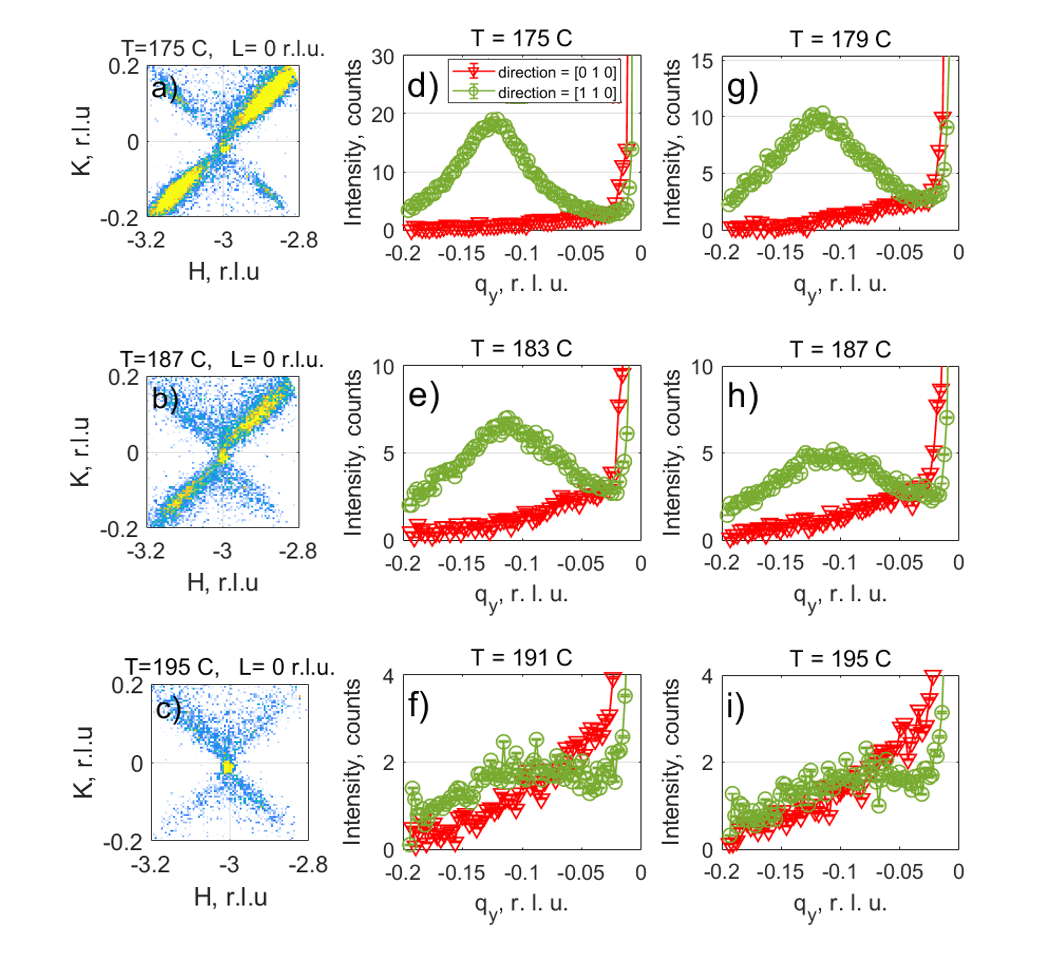}
\caption{\label{DS_IM}
Diffuse scattering distributions in the IM phase. Two-dimensional distributions (cuts througth $Q$ = (-3 0 0) by $H$ $K$ $0$ plane) are shown using same color scale. 
The profiles correspond to [1 1 0] (circles) and [0 1 0] (triangles) directions, background subtracted. 
}
\end{figure}

\begin{figure}
\includegraphics[width = 0.7\columnwidth]{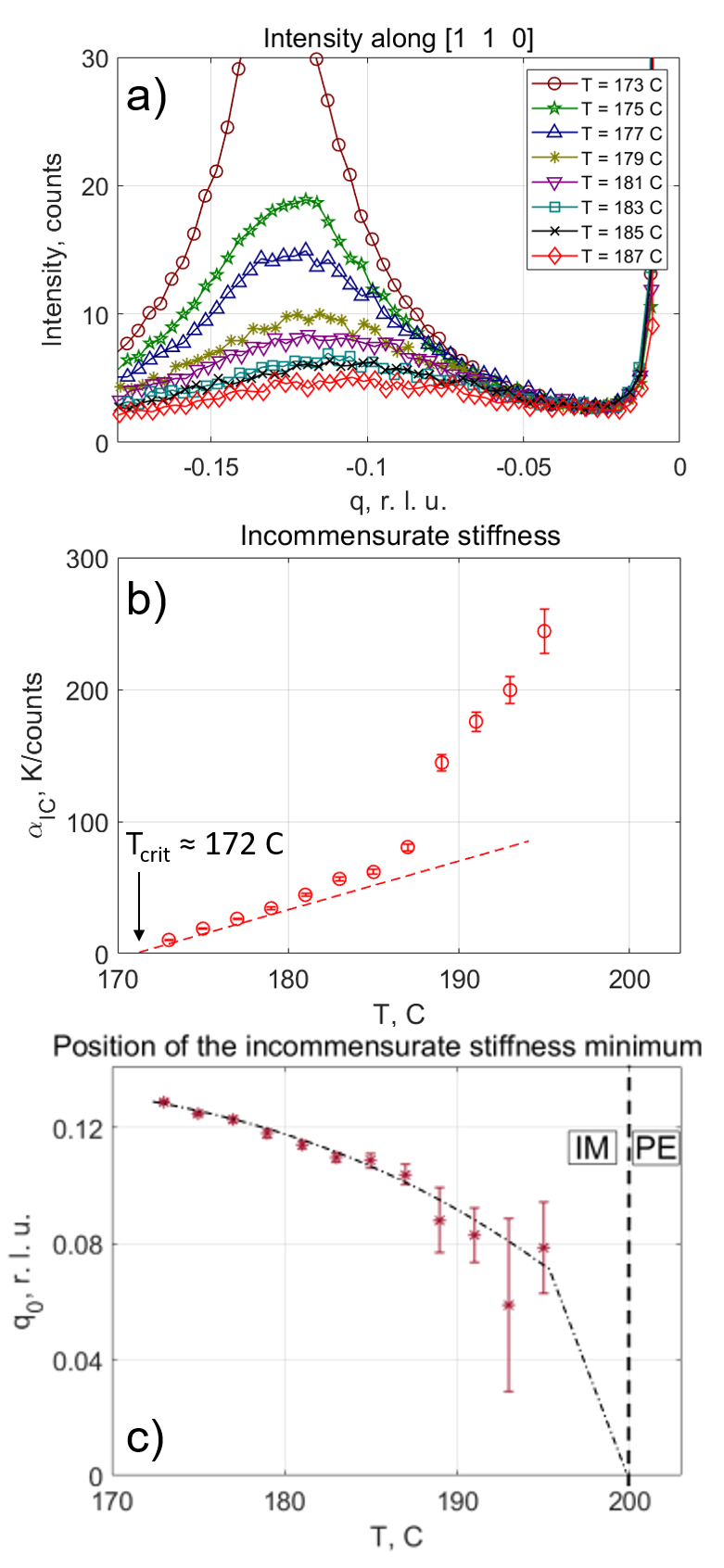}
\caption{\label{alpha_IM}
Temperature dependence of the incommensurate diffuse scattering in the IM phase (a) and its parameters: incommensurate stiffness, measured as T/I$_{max}$ (b), and  position of intensity maximum (c).
}
\end{figure}

Since the transition from the IM phase to the AFE2 phase is incommensurate, it is logical to expect  a decrease in the stiffness corresponding to transverse incommensurate waves. Indeed, we register a minimum of stiffness $\alpha_\text{IC}$ at an incommensurate position, $q_0$. This can be seen from the scattering profiles [Figs. \ref{DS_IM}, \ref{alpha_IM}(a)], where the maximum is located in the vicinity of $q_0$ = (0.15, 0.15, 0). The position of the maximum shifts away from the center of the zone as sample cools down [Fig. \ref{alpha_IM} (c)].

The most interesting feature of this maximum is that the incommensurate stiffness, recalculated from its amplitude, behaves linearly when approaching the IM $\rightarrow$ AFE2 transition and the critical temperature for this linear dependence coincides with the transition temperature [Fig. \ref{alpha_IM} (b)]. At the same time, the position of the minimum of the stiffness just above the transition coincides with the position of the incommensurate  reflection just below it. These two factors are consistent with what would be expected from classical soft-mode-driven (second order) incommensurate transition \cite{BlincLevanyuk}. However, this apparent similarity to second-order transition does not directly agree with the presence of non-zero latent heat in caloric measurements of such crystals \cite{Sumara2017}. This discrepancy is of interest, although no immediate explanation is apparent, apart from a possibility that the experimentally registered latent heat is due to extrinsic sources, such as domain redistribution at the transition point.

The linear trend of incommensurate stiffness breaks at  $T = 188^{\circ}$ $C$, where the slope changes (Fig. \ref{alpha_IM} (a)). Seemingly, this is due to a change in the distribution of IM-phase domains with different orientations at $T$ about $187 - 189^{\circ}$ $C$, as Fig. \ref{dom_pop} indicates.

\section{Interpretation}

Here, a symmetry-guided minimal Landau-like model is built and the experimental data are compared to its predictions.

\subsection{Symmetry of the IM phase}

The structure of IM phase has not yet been refined crystallographically. However, plausible assumptions on it can be made on the basis of its AFD character, general symmetry considerations and experimentally observed symmetry of diffuse scattering. 

The first candidate for IM-phase unit cell is the orthorhombic one defined by vectors $\vec{a}_\text{IM} = (a_\text{PC}, -a_\text{PC}, 0)$, $\vec{b}_\text{IM} = (a_\text{PC}, a_\text{PC}, 0)$, $\vec{c}_\text{IM} = (0,0, 2a_\text{PC})$, where $a_\text{PC}$ is the pseudocubic lattice parameter. Oxygen octahedra are tilted in $a^-a^-c^0$ pattern around $\vec{a}_\text{IM}$. Such a cell of \textit{Imma} symmetry has been considered by Toledano and Khalyavin \cite{Toledano2019} (axes $a$, $b$ and $c$ assigned differently in the present work to have them similar to those in the AFE phase) as a virtual intermediate structure between cubic and antiferroelectric structures. Also, this cell corresponds to the average structure of the incommensurate phase of PbHfO$_3$, as determined crystallographically \cite{Bosak2020}.

\begin{figure}
\includegraphics[width = \columnwidth]{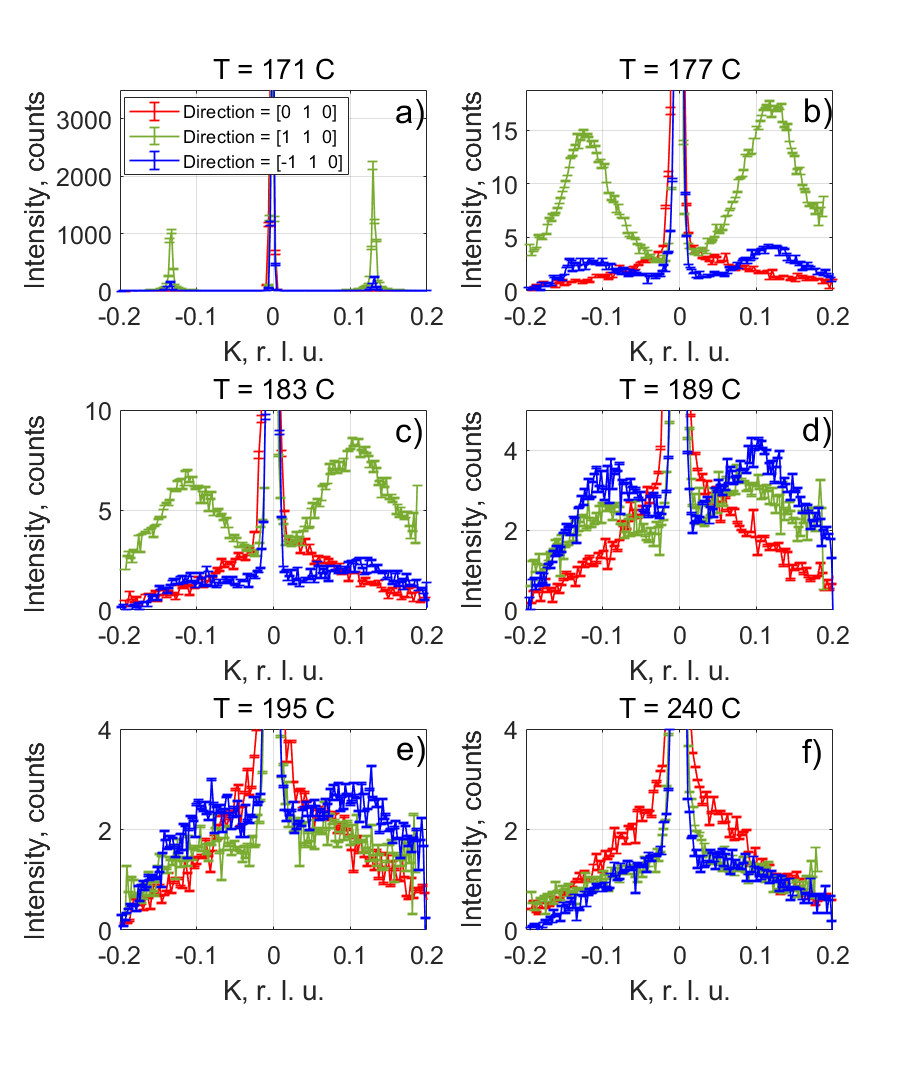}
\caption{\label{dom_pop}
Population of domains in the AFE2 and IM phases. In the AFE2 phase (a), as well as in the low-temperature region of the IM phase (b, c), domains with incommensurate maxima along the [1 1 0] direction strongly prevail in volume over domains with incommensurate maxima along the [-1 1 0] direction, and other directions (not shown). However, as the transition to the cubic phase is approached, the intensities along the [1 1 0] and [-1 1 0] directions are equalized (d, e). Cubic phase (f) shows equal intensities in these directions.
}
\end{figure}

The orthorhombic \textit{Imma} cell is compatible with the observed symmetry of diffuse scattering. 
This can be deduced as follows. If one considers a temperature near the bottom of IM phase stability range, like $T=177^\circ$ C, the diffuse scattering maxima along [1 1 0] direction are 5-7 times more intense than those along [1 -1 0], [1 0 1] and [1 0 -1]. Intensity profiles along [1 1 0] and [1 -1 0] are shown in Fig. \ref{dom_pop} (b), while the latter two directions ([1 0 1] and [1 0 -1]) are omitted for clarity of the figure, intensity of respective profiles is about the same as along [1 -1 0] at this temperature. It is possible to verify that such a combination of intensities cannot be reproduced by any combinations of rhombohedral-symmetry domains or tetragonal-symmetry domains, but is compatible with orthorhombic-symmetry domains. 
The volume shares of different orientational domain states are changing with temperature, as commented in the caption of Fig. \ref{dom_pop}.


\subsection{Minimal Landau-like model of the IM phase}

By knowing (actually -- reliably guessing) the order parameter of IM phase, it is possible to learn how the stiffness landscape in wavevector space should be organized. Order parameter is assumed to be $\vec{\eta}_0 =$ $\eta_0$($\frac{1}{\sqrt{2}}, \frac{-1}{\sqrt{2}}, 0$): anti-phase tilts around $\vec{a}_\text{IM}$. 
The idea is to build such a symmetry-determined free energy expansion, in which by means of the lowest-possible-order terms involving $\vec{\eta}_0$ and polarization gradients, the key specifics of the IM phase are accounted for. 

\subsubsection{Cubic part of the free energy expansion}
Free energy of inhomogeneous polarization in a cubic crystal is given as \cite{Yudin2014} 
\begin{equation}\label{eq_cubic_polarization}
    F_P = \frac{\alpha}{2} \vec{P}^2 + \frac{1}{2}\sum_{i,j} D_{ijlm} \frac{ \partial P_i }{\partial x_i} \frac{ \partial P_l }{\partial x_m},
\end{equation}
where $\vec{P}$ -- polarization, $\alpha$ -- dielectric stiffness, $D_{ijlm}$ -- polarization correlation energy tensor, $x_i$ -- coordinates $x$,$y$ and $z$. This formula does not account for the LO-TO splitting due to the long-range electrostatic interactions, which should be, in principle, included in such macroscopic consideration \cite{Marton2006simulation}. Here, for simplicity, we do not consider them, but instead (which is largely equivalent) just assume that all possible polarization waves in Eq. \eqref{eq_cubic_polarization} are transverse.


\subsubsection{Interaction between polarization and tilts}
Interaction of polarization with $\vec{\eta}_0$ should be described by those coupling terms between $\Gamma_4^-$ (for polarization) and $R_4^+$ (for tilts) distortions that the cubic symmetry allows. Supplemental Material \footnote{See Supplemental Material at [URL will be inserted by publisher] for ISOTROPY software output describing the symmetry-allowed terms of free energy expansion in terms of polarization and tilts. } lists such terms up to fourth order (and up to second order in gradients), as provided by ISOTROPY software \cite{ISOTROPY}. 

The commonly considered coupling is the bi-quadratic coupling \cite{Holakovsky1973,Tagantsev2013}, which is represented by three fourth-order terms. In our case they are proportional, respectively, to
\begin{equation}\label{eq_bi_1}
    (P_x^2 + P_y^2)\eta_{0}^2 \text{, } 
    P_x P_y\eta_{0}^2 \text{ and }
    P_z^2\eta_{0}^2.
\end{equation}
This bi-quadratic coupling between $\vec{\eta}_0$ and homogeneous polarization accounts for the change in dielectric tensor between cubic and IM phases, but does not account for the change in spatial correlations of polarization.

\subsubsection{Bi-quadratic coupling of tilts to polarization gradient}
Spatial correlations should be affected by coupling terms involving polarization gradients. Also those terms should not contain $\vec{\eta}_0$ gradients, because those should be zero for constant $\vec{\eta}_0$. 
Lowest order terms of this sort are degree-3 terms linear in polarization gradient, which vanish in a big crystal, because their volume integrals are transformed to surface integrals of finite vector fields \cite{Landau2013statistical}. Therefore, the desired lowest-order interaction should be described by bi-quadratic terms between polarization gradients  and homogeneous tilts, of which there are 30 invariants (only 15 are non-equivalent in big crystals) with terms proportional to \cite{Note1}
\begin{equation}
    \frac{\partial P_i}{\partial x_j}
    \frac{\partial P_l}{\partial x_m} 
    \eta_s\eta_t.
\end{equation}
Upon substituting $\vec{\eta}_0 =$ $\eta_0$($\frac{1}{\sqrt{2}}, \frac{-1}{\sqrt{2}}, 0$), these terms contribute to free energy as 
\begin{equation}
    \Delta F = 
    \frac{1}{2}
    \eta_0^2 
    \sum_{i,j} \gamma_{ijlm} \frac{ \partial P_i }{\partial x_i} \frac{ \partial P_l }{\partial x_m}.
\end{equation}
Symmetry of $\gamma_{ijlm}$ tensor is not cubic, as that of $D_{ijlm}$, but orthorhombic with main axes along $\vec{a}_\text{IM}$ ([-1 1 0]), $\vec{b}_\text{IM}$ ([1 1 0]) and $\vec{c}_\text{IM}$ ([0 0 1]).
%
The coefficient $\gamma_{abab}$, which determines the dispersion of transverse polarization waves propagating along $\vec{b}_\text{IM}$ and polarized along $\vec{a}_\text{IM}$ reads
\begin{equation}
    \gamma_{abab} = \gamma_{xxxy} - \gamma_{xyyy} + \frac{1}{2} (\gamma_{xxxx} - \gamma_{xxyy} + \gamma_{xyxy} - \gamma_{xyyx}).
\end{equation}
This coefficient is different from $\gamma_{baba}$, which corresponds to similar waves, but propagating along $\vec{b}_\text{IM}$. Therefore, bi-quadratic coupling between polarization gradient and tilts can be sufficient for creating the incommensurate stiffness minimuma selectively at $\vec{q}_0 = \pm(\xi,\xi,0)$, without simultaneously creating such minima at all the wavevectors of the same star. 


Correlation energy for transverse waves propagating along $\vec{b}_\text{IM}$ is defined by constant $D_2$, which is recalculated from tensors as
\begin{equation}\label{D2_renormalized}
    D_2 = (D_{xxxx} - D_{xxyy}) + \gamma_{abab}\eta_0^2.
\end{equation}
For an incommensurate stiffness minimum this coefficient shall be less than zero.

\subsubsection{Expected temperature dependences of coefficients}
Following the general philosophy of Landau-like models, such as Devonshire's \cite{Devonshire1949}, one tries to find a minimal number of parameters that depend on temperature, ideally -- one parameter, which is linear in $T$, as in Landau theory \cite{Levanyuk2020}. Here, one needs at least two parameters, because in the cubic phase both the ferroelectric and AFD modes appear soft and the model does not present a natural mechanism of linking one softening to the other. Therefore, the simplest model would be with both dielectric and AFD stiffness following linear trends, 
\begin{equation}
    \alpha(T) = A(T-T_0)
\end{equation} 
and 
\begin{equation}
    \alpha^\text{AFD}(T) = A^\text{AFD}(T-T_0^\text{AFD}).
\end{equation}

\begin{figure}
\includegraphics[width=.9\columnwidth]{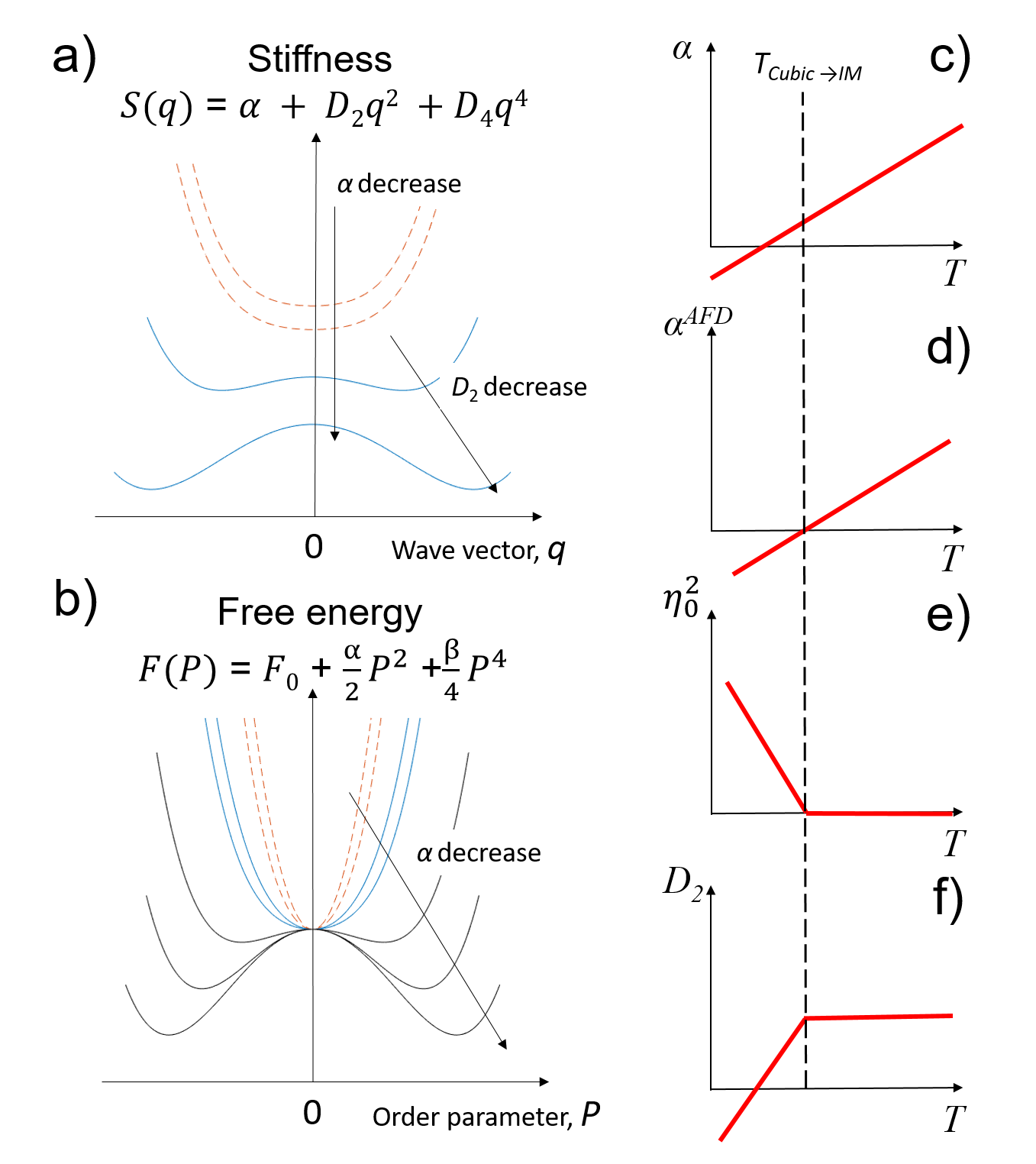}
\caption{\label{Landau-like}
Minimal Landau-like model of IM phase in temperature domain. Ferroelectric stiffness, $\alpha$, decreases linearly across the cubic$\rightarrow$IM transition. In the cubic phase (dashed red lines) this leads to homogeneous shift of the whole stiffness dispersion in (a) and to unbending of free energy parabolic part in (b). In the IM phase (blue solid lines), the polarization correlation energy, $D_2$, linearly decreases due to the bi-quadratic interaction of polarization gradient with oxygen octahedral tilts. This leads to the unbending of stiffness dispersion in (a) and formation of incommensurate stiffness minimum. Subsets (c) - (f) describe temperature behaviour of the model parameters: dielectric stiffness $\alpha$, AFD stiffness $\alpha^{AFD}$, squared AFD order parameter $\eta_0^2$ and correlation energy constant $D_2$, respectively.
}

\end{figure}

Landau theory does not allow \textit{Pm}$\bar{3}$\textit{m}$\rightarrow$\textit{Imma} AFD transition to be of second order \cite{Howard1998group}. On the other hand, the temperature dependence of $R$-point reflections appear linear in the IM phase (Fig. \ref{Phases}), which would be consistent with second-order transition. The simplest reconciliation is to assume that the transition is of ``\textit{weakly}'' first order and $\eta_0^2$ is not strictly, but \textit{nearly} linear just below cubic$\rightarrow$IM transition, 
\begin{equation}
    \eta_0^2 \approx Y(T_0^\text{AFD}-T)
\end{equation}
(above $\approx T_0^\text{AFD}$, $\eta_0^2=0$). 
The polarization correlation energy constant will then be also nearly linear (Eq. \eqref{D2_renormalized}) and decrease on cooling if $\gamma_{abab}$ is negative. 
Fig. \ref{Landau-like} shows how $q$-dependent stiffness evolves in this case. Note that this behavior is ultimately simplified: along with the assumptions above, bi-quadratic coupling to homogeneous polarization (as opposed to the coupling to gradients) is neglected. Specifics of this minimal model is that the stiffness minimum shifts from zero in wavevector space nearly continuously, similarly to how the free energy minimum shifts from zero along the polarization axis in Landau theory of second-order transitions.

\subsection{Model vs. experiment}
\begin{figure*}
\includegraphics[width=0.9\textwidth]{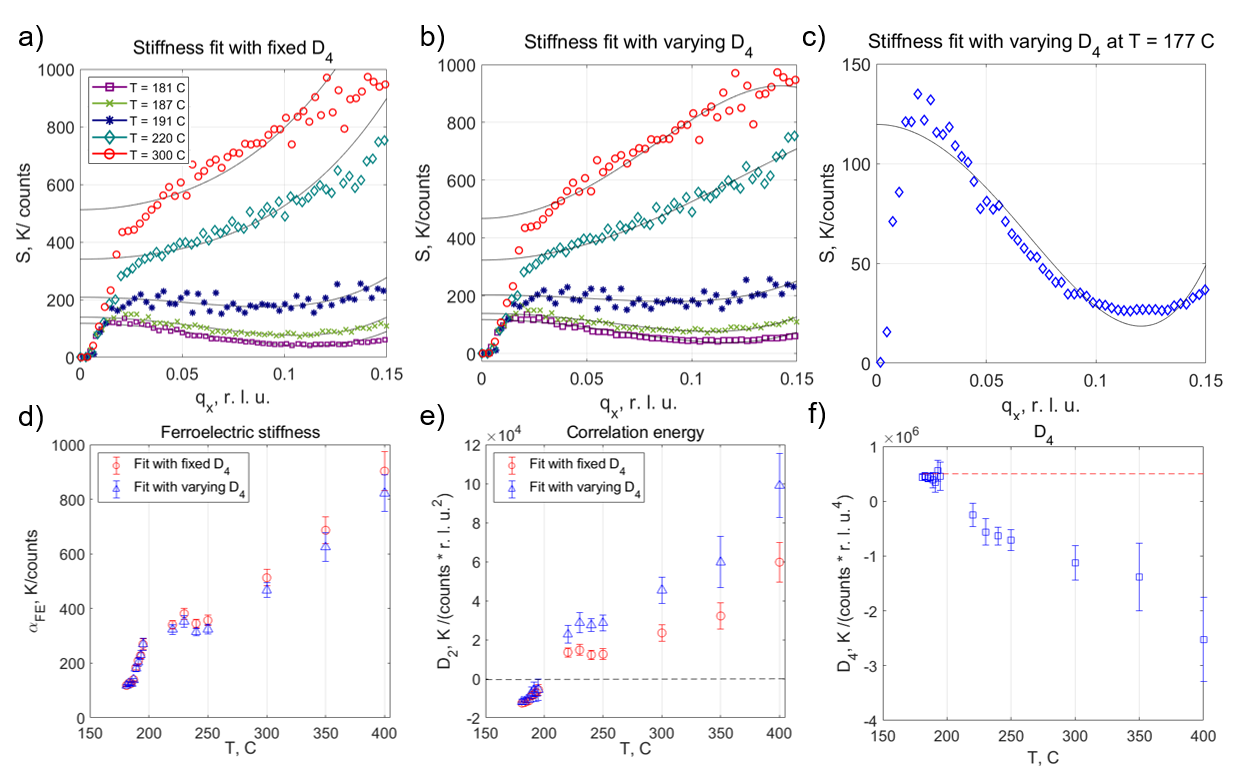}
\caption{\label{ModelToExp}
Description of the stiffness profiles along the [1 1 0] direction by the relation $S(q) = \alpha_\text{FE} + D_2q^2 + D_4q^4$, where $\alpha_\text{FE}$ is the ferroelectric stiffness and $D_2$ is the polarization correlation energy constant. At low $q_x$ values, the Bragg scattering and scattering by acoustic phonons are dominant, so stiffness fit is done for 0.03 $<$ $q_x$ $<$ 0.15 r.l.u. The $\alpha_\text{FE}$ and $D_2$ values are presented in non-SI units; to switch to SI units, it is necessary to multiply the absolute values of the parameters by $N_1 = 1.23 \times 10^5 [Jm/C^2]$ (for the stiffness) and $N_2 = 2.05 \times 10^{-14} [m^3/C^2]$ (for the $D_2$), respectively. The stiffness in this figure is the same in meaning as in Fig. \ref{alpha_FE}, but slightly differs in value due to different methods of obtaining it: Fig. \ref{alpha_FE} obtained from the fit of the intensity (proportional to the susceptibility) taking into account only $D_2$, and here it was done from the fit of the stiffness, taking into account also $D_4$.
}

\end{figure*}

This section examines the experimental data in comparison with the minimal model outlined above. There are no attempts to fit the data using model temperature dependences (they are considerably different), but rather the $q$-dependent stiffness is modeled by respective formula $S(q) = \alpha_\text{FE} + D_2q^2 + D_4q^4$ [Fig. \ref{ModelToExp} (a-c)] at each temperature point and the resulting temperature dependences of coefficients [Fig. \ref{ModelToExp} (d-e)] are compared with expectations [Fig. \ref{Landau-like} (c,f)]. The fitting result with temperature-independent $D_4$, in our opinion, differs in shape with  experiment in the high-temperature region of the cubic phase. So, we tried also a temperature-dependent $D_4$, which agrees better. 
The profiles in Fig. \ref{ModelToExp} (a,b) correspond to $T > 181^{\circ} \text{C}$. The fits of these profiles appear reasonable. On the other hand, at lower temperatures ($173^{\circ} \text{C} < T < 179^{\circ} \text{C}$) the fits appear inadequate, as Fig. \ref{ModelToExp} (c) illustrates: the shape is clearly different. 

Both  $\alpha$ and  $D_2$ decrease on cooling in the cubic and IM phases. This is especially remarkable for $D_2$: while it is expected to change in the IM phase due to interaction with tilts [Eq. \eqref{D2_renormalized}, Fig. \ref{Landau-like}(f)], in the cubic phase it is not expected to change. In Landau theory it is temperature-independent \cite{Landau2013statistical}, and one could expect this to extend qualitatively to Landau-like models. For the present crystal this seems not applying, likely because it is not so similar to ferroelectrics, at least its $\alpha(T)$ does considerably violate $\alpha = A(T-T_0)$ law and no ferroelectric phase forms. 

\begin{figure}
\includegraphics[width = 1\columnwidth]{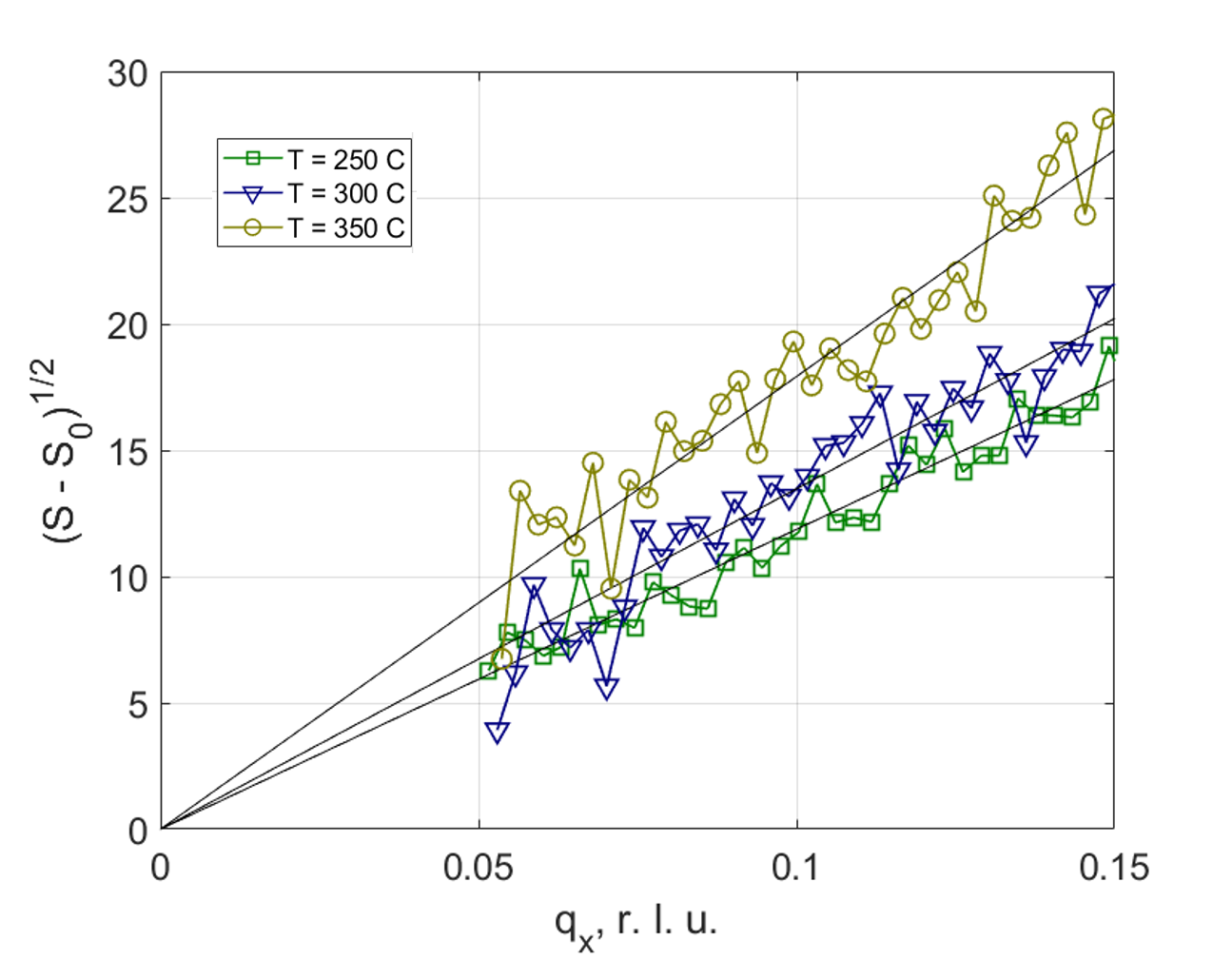}
\caption{\label{S_0_sqrt}
The square root of the $q$-dependent part of the stiffness profiles along the [1 1 0] direction in the cubic phase. Slopes correspond to $\sqrt{D_2}$.
}

\end{figure}

Upon noting the interesting temperature dependence of $D_2$ in the fits, it is instructive to attempt ensuring that this effect is real and not an artefact of the data processing. Fig. \ref{S_0_sqrt} supports the fit results by showing stiffness profiles, recomputed so that the $y$-axis becomes $\sqrt(S-S_0)$, where $S$ is the stiffness and $S_0$ is the stiffness at the zone centre, as obtained by the fit. This way the parabolic stiffness curve is transformed to a linear curve with the slope of $\sqrt(D_2)$. This slope increases with temperature.

\section{Discussion}

\subsection{Relation to other observations of intermediate phase}

The present IM phase is similar to the additional phase in pure PbHfO$_3$ that is stabilized under pressure, where similar diffuse incommensurate maxima, as well as the $R$-point reflections, were also registered \cite{Knyazeva2019}. There are subtle differences, however, in part due to the presence of weak reflections at  $X$- and $M$-points under pressure, which IM phase in PbHfO$_3$-PbSnO$_3$ lacks. Similar incommensurate maxima were also seen in PbZrO$_3$ under pressure \cite{Burkovsky2017}, although, unfortunately, without information on the $R$-point superstructures. Most likely, all these phases are similar. 
Other studies of PbHfO$_3$-PbSnO$_3$ \cite{jankowska2020local, jankowska2021complexity}, where PbSnO$_3$ concentration was lower, did not reveal such a temperature-dependent incommensurate diffuse scattering maxima. Tentatively, this behavior changes strongly with tin concentration.

\begin{figure}
\includegraphics[width = 1\columnwidth]{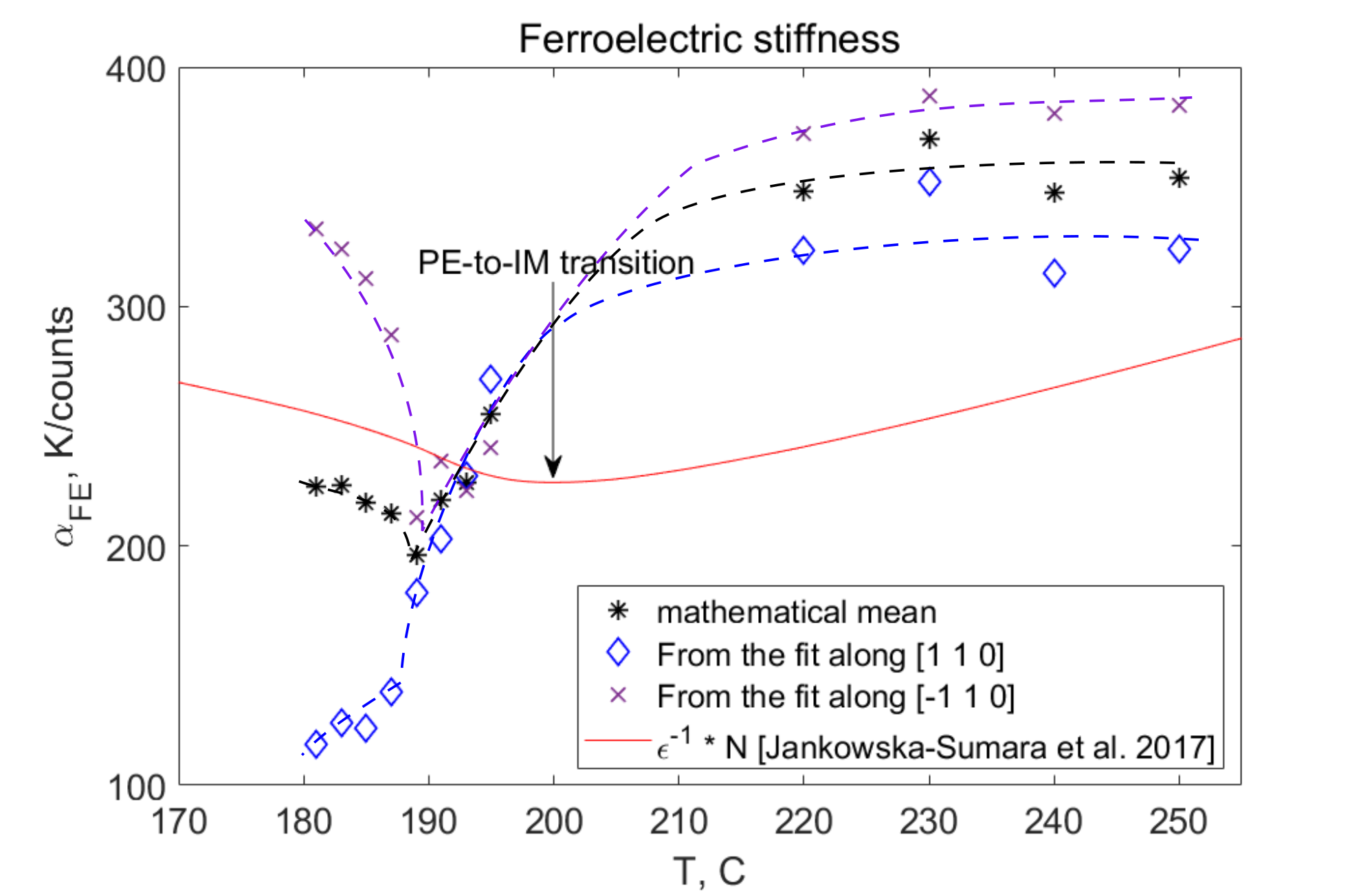}
\caption{\label{Stiff_lines}
Ferroelectric stiffness values extracted from diffuse scattering profiles along the [1 1 0] and [-1 1 0] directions. Average of those values is compared with the dielectric data of Jankowska-Sumara \textit{et al.} \cite{Sumara2017}. One notes that averaging over two different directions results in better agreement in temperature trends below the transition point than the stand-alone values related to different directions. 
}

\end{figure}

There is a disagreement between the temperature trend of the IM-phase dielectric stiffness obtained by diffraction [see Fig. \ref{ModelToExp}(d)] and the corresponding trend obtained by dielectric measurements \cite{Sumara2017}. From diffraction, dielectric stiffness should decrease on cooling, while from dielectric measurements it should increase. We think this discrepancy comes likely from poly-domain structure
of the IM phase. The agreement tends to recover (Fig. \ref{Stiff_lines}), if one considers a stiffness produced by averaging diffuse-scattering-derived values related to displacements along different \{110\} directions, which is more reminiscent to what is measured in a dielectric setup.

\subsection{Inhomogeneous polarization is not sufficient for modeling incommensurate fluctuations in the lower-$T$ part of the IM phase}

It is remarkable that the minimal symmetry-based Landau-like model describes the data not so well, as the comparison of Figs. \ref{Landau-like} and \ref{ModelToExp} highlights. The discrepancies between model and experiment [Fig. \ref{ModelToExp}(c)] arise surprisingly quickly, in about 20$^{\circ}$C below the cubic$\rightarrow$IM transition. Normally, one should expect the validity region for Landau-like models of ferroelectrics to be larger, which is related to the experimental temperatures being small when compared to the so-called ``atomic temperature'' of about $10^4$ -- $10^5$ K \cite{Levanyuk2020}. Therefore, the origin of these discrepancies is presently unclear. 

What one may suggest from the discrepancy of Fig. \ref{ModelToExp}(c) is that the incommensurate order parameter, when its wavevector is considerably away from the zone center, cannot be interpreted as just an inhomogeneous polarization. It is not necessary due to stiffness renormalization by flexoelectric or similar interactions, but may be simply because sufficiently away from the zone center the waves cannot be considered as \textit{weakly} inhomogeneous.
An independent order parameter should be introduced instead, for which the symmetry-allowed expansion around its wavevector, $\vec{q}_0$, is expected to be adequate in the vicinity of IM$\rightarrow$incommensurate transition. An experimental indication towards this is the linear dependence of incommensurate stiffness near that transition [Fig. \ref{alpha_IM} (b)], which is compatible with classical models of incommensurate transitions with explicitly defined incommensurate order parameter \cite{BlincLevanyuk}. If, instead, one assumes that incommensurate stiffness follows ferroelectric softening, as in the minimal model considered here, the incommensurate stiffness would be quadratic in temperature, which contradicts the observations. It seems appropriate to suggest that the dielectric stiffness may be not very successful candidate for being a critical parameter in models like this, despite the experimental increase of $\epsilon(T)$ on cooling in the cubic phase.

\subsection{Minimal model with gradient terms as a generalization of Holakovsky's model of triggered ferroelectricity}

Relation of dipole ordering to the tilts subsystem in perovskite and similar structures is widely discussed \cite{Holakovsky1973, Zhong1995Competing, Tagantsev2013, Benedek2011hybrid, Bellaiche2013, Iniguez2014, Mani2015, Burkovsky2018, Gu2018, Burkovsky2019, Xu2019order,Schranz2020contributions}. The likely initial point of that discussion is Holakovsky’s proposition \cite{Holakovsky1973} on a possibility that it is the octahedral tilt subsystem that can be ``soft'' instead of ``ferroelectric'' subsystem, while the ferroelectric transition still taking place due to a positive bi-quadratic interaction between those. The present paper attempts extending this to inhomogeneous polarization by considering bi-quadratic coupling also to polarization gradients. This appears successful in the higher-$T$ region of IM phase, where incommensurate fluctuations occur near the zone center, but fails at low-$T$ region, where the fluctuations are of much shorter wavelength ($q \approx (1/7, 1/7, 0)$. In pure PbHfO$_3$, the region of an apparent validity of the minimal model is skipped, because the IM phase itself is skipped, and the incommensurate modulations are of relatively short wavelength from the very beginning. The minimal model is, therefore, unlikely sufficient (although it is conceptually consistent) for describing triggered incommensurate transition in pure PbHfO$_3$ \cite{Burkovsky2019} and a more complicated model is needed.

\subsection{A possibility of temperature-independent characteristic length scale for inhomogeneities}

Polarization correlation energy constant, $D_2$, appears to follow a similar temperature trend to the one of ferroelectric stiffness: they both decrease nearly linearly in the cubic phase with critical temperature of about 200 $^\circ$ C. If one composes the correlation length parameter from those as $\sqrt{D_2/\alpha}$, this parameter appears rather temperature-independent. If so, one should think that this length scale is characteristic to the system in a wide temperature range. Upon estimating this length scale from Fig. \ref{ModelToExp} (d,e), one arrives at the value of about 10 lattice units (projected to pseudocubic high-symmetry directions). This value is comparable to the period of incommensurate phase, which is $\approx 7$ lattice units. A natural suggestion seems to be that the fluctuations of polarization, even in the cubic phase, are on the average as inhomogeneous as the incommensurate phase is, although the level of order differs drastically -- from long-range order in the incommensurate phase through short-range correlated incommensurate fluctuations in the IM phase and to, apparently, much more disordered fluctuations in the cubic phase. This is similar to how the local structure of some perovskites appears largely temperature-independent despite drastic changes in the macroscopic symmetry and structure \cite{Teslic1998,Jiang2013local}.



\section{Conclusion}

This work brings a useful perspective on the functioning mechanisms of lead-based antiferroelectrics and methodology of their scattering studies. In particular, we have identified linear temperature dependence for incommensurate soft mode stiffness, a crossover from ferroelectric-like to incommensurate fluctuations upon cubic$\rightarrow$IM phase transition, unusual indication towards largely temperature-independent length scale of inhomogeneities. The observations were tested against a minimal symmetry-constrained Landau-like model, which revealed a rather limited applicability range of such models to such crystals.


\begin{acknowledgments}
N. Ter-Oganessian and Peng Chen are acknowledged for useful discussions on symmetry-related subjects. M. Polentarutti and G. Bais are acknowledged for useful discussions and proofreading. A. Bosak is acknowledged for his help with experiment and useful discussions.

The reported study was supported by Russian Foundation for Basic Research, project numbers 20-32-90176 and 20-32-70215. The authors acknowledge the European Synchrotron Radiation Facility for provision of synchrotron radiation facilities.
\end{acknowledgments}

\section*{Appendix}

\subsection{Structure factor for diffuse scattering by a simple lattice}

In the approximation of small displacements, $\vec{U}_i$, from high-symmetry positions, $\vec{R}_i$, the total structure factor for a simple lattice can be expanded as follows (Debye-Waller factor and normalization factors are omitted for simplicity).

\begin{equation}
\begin{gathered}
{F}(\vec{Q}) = 
{f}(\vec{Q}) \sum\limits_{i}{\text{exp}(i\vec{Q}\vec{r_i})} =
\\
{f}(\vec{Q}) \sum\limits_{i}{\text{exp}(i\vec{Q}(\vec{R_i} + \vec{U_i}))} \approx 
\\
{f}(\vec{Q}) \sum\limits_{i}{\text{exp}(i\vec{Q}\vec{R_i})(1+i\vec{Q}\vec{U_i})} =
\\
{f}(\vec{Q})\delta(\vec{Q} - \vec{\tau}) + i{f}(\vec{Q})(\vec{Q}\vec{U}_q)\delta(\vec{Q} - \vec{q} - \vec{\tau}),
\end{gathered}
\end{equation}
where $f(\vec{Q})$ is atomic scattering factor, $\vec{U}_q$ -- Fourier component of displacements. In the last row, the former term corresponds to a contribution of this sublattice to Bragg scattering structure factor, the latter -- to structure factor for diffuse scattering.





\bibliography{references,bibliography_GoldenAutumn,bibliography_WarmWinter}

\end{document}